\documentclass{article}
\usepackage{dcase2021,amsmath,graphicx,url,times,booktabs, tabularx}

\usepackage{color, colortbl}
\definecolor{Gray}{gray}{0.9}

\usepackage{siunitx, gensymb, adjustbox}

\title{STARSS22: A dataset of spatial recordings of real scenes with spatiotemporal annotations of sound events}

\name{Archontis Politis$^{1}$,
      Kazuki Shimada$^{2}$,
      Parthasaarathy Sudarsanam$^{1}$,
      Sharath Adavanne$^{1}$,
      Daniel Krause$^{1}$
      }
\secondlinename{
      Yuichiro Koyama$^{2}$,
      Naoya Takahashi$^{2}$,
      Shusuke Takahashi$^{2}$,
      Yuki Mitsufuji$^{2}$
      Tuomas Virtanen$^{1}$,
      }
\address{$^1$ Audio Research Group, Tampere University, Tampere, Finland\\          
        $^2$ Sony Group Corporation, Tokyo, Japan
 }

\begin{document}

\ninept
\maketitle

\begin{sloppy}

\begin{abstract}
This report presents the Sony-TAu Realistic Spatial Soundscapes 2022 (STARS22) dataset for sound event localization and detection, comprised of spatial recordings of real scenes collected in various interiors of two different sites. The dataset is captured with a high resolution spherical microphone array and delivered in two 4-channel formats, first-order Ambisonics and tetrahedral microphone array. Sound events in the dataset belonging to 13 target sound classes are annotated both temporally and spatially through a combination of human annotation and optical tracking. The dataset serves as the development and evaluation dataset for the Task 3 of the DCASE2022 Challenge on Sound Event Localization and Detection and introduces significant new challenges for the task compared to the previous iterations, which were based on synthetic spatialized sound scene recordings. Dataset specifications are detailed including recording and annotation process, target classes and their presence, and details on the development and evaluation splits. Additionally, the report presents the baseline system that accompanies the dataset in the challenge with emphasis on the differences with the baseline of the previous iterations; namely, introduction of the multi-ACCDOA representation to handle multiple simultaneous occurences of events of the same class, and support for additional improved input features for the microphone array format. Results of the baseline indicate that with a suitable training strategy a reasonable detection and localization performance can be achieved on real sound scene recordings. The dataset is available in \url{https://zenodo.org/record/6387880}.
\end{abstract}


\section{Introduction}
\label{sec:intro}

Sound event localization and detection (SELD) refers to the task attempted by methods aiming to simultaneously detect the presence and track the location or direction of certain sound types of interest in a sound scene over time. The task relates strongly to the more established ones of sound event detection (SED) and sound source localization (SSL) but it adds spatial information to the first and semantic information to the second and, hence, it opens some further possibilities on machine listening, robot audition, acoustic monitoring, and human-machine communication, among others.

The SELD task has recently gained interest and popularity in the audio research community, in part due to its introduction in the DCASE Challenge in 2019 which gave the opportunity to researchers to test and compare their methods on a standardized dataset and against a common baseline. The challenge submissions were analyzed and discussed in the overview of \cite{politis2020overview}. The dataset of the challenge was generated with a collection of spatial room impulse responses (SRIRs) from 5 spaces and multiple source positions convolved with dry isolated sound event recordings \cite{Adavanne2019a}. The next iteration of DCASE2020 increased the diversity of the training and testing conditions by including SRIRs of 10 additional rooms with stronger reverberation and, more importantly, by emulating dynamic scenes with both moving and static sound sources \cite{Politis2020}. The same sound scene generation process was used in the third iteration of the challenge in DCASE2021 increasing however scene complexity by adding directional interfering events out of the target classes \cite{Politis2021}.

The three previous SELD challenges contributed to the continuous development and improvement of SELD methods by taking care to emulate faithfully the spatial and acoustical properties of sound scenes and to gradually increase scene complexity, bringing every iteration closer to real conditions. However, there are certain limitations inherent to generating synthetic mixtures that have persisted through the previous iterations. An example of such limitations is the random presence of target classes in a scene and the random sequencing of sound events, discarding the natural temporal occurrences or co-occurrences of certain sounds happening in a real scene. Another such limitation is the randomized spatial distribution of sound events ignoring the fact that many events result from the actions of certain agents in a scene and are spatially connected. Hence, to overcome those limitations SELD systems should transition to training and evaluation with recordings of real sound scenes. Datasets of real sound scenes require human annotation, a hard task even in the case of SED only, while the task of simultaneous spatial annotations requires some form of automated tracking as it would be impossible to be performed by humans. Due to this complexity, there are no published SELD datasets we know of except for the SECL-UMons one in \cite{brousmiche2020secl}, capturing natural sound events of 11 classes in two spaces, resulting from actions at pre-defined locations in each room. However, it consists of recordings of a single such event in isolation or combinations of two simultaneous events, and even though it contains events at natural spatial distributions it ignores the variability and diversity of sounds in a natural scene with multiple agents being linked both temporally and spatially. A few more synthetic SELD datasets exist with the same limitations as the previous DCASE datasets, based on captured SRIRs and targeting certain applications, such as wearable arrays \cite{nagatomo2022wearable} or positional localization in a room with distributed arrays \cite{guizzo2021l3das21}.

This report presents the first SELD dataset we are aware of where natural scenes, loosely acted by multiple actors, are captured and annotated with strong labels temporally and spatially. The challenges of such annotations are dealt with a combination of human listening and optical tracking, employing multiple sensors and modalities. Since the sound scenes are acted naturally, the dataset overcomes the limitations of synthetic datasets discussed earlier. Target sound classes do not follow some random combination but are instead constrained by the environment and the participants, while the presence of each class is determined by the natural composition of each scene. Causal and sequential occurrences of sound events, as well as co-occurrences, follow the actions of the actors and their interactions with the environment. The same occurs with the location of events and their trajectories in case they are moving; their spatial distributions are naturally constrained by the type of event, while event trajectories can reveal scene information on the agents and their actions. Hence, the dataset opens certain new possibilities for SELD systems, apart from allowing evaluation in realistic scenarios. 

The STARSS22 dataset serves as the development and evaluation dataset of DCASE2022 Task 3, and it is followed by a suitable baseline and evaluation setup. Changes in the baseline or the evaluation setup with respect to the previous DCASE challenges are elaborated. Since the duration of the dataset is limited compared to the synthetic datasets used in previous years, the use of external data is allowed in this iteration to improve model training and generalization. An example strategy based on additional synthetic data is presented for the baseline. Finally, results are presented on the development set showing reasonable SELD performance.

\section{Dataset}
\label{sec:dataset}

The \textbf{Sony-TAu Realistic Spatial Soundscapes 2022}(\textbf{STARSS22}) dataset consists of recordings of real scenes captured with high channel-count spherical microphone array (SMA). The recordings are conducted from two different teams at two different sites, Tampere University in Tampere, Finland, and Sony facilities in Tokyo, Japan. Recordings at both sites share the same capturing and annotation process, and a similar organization. They are organized in sessions, corresponding to distinct rooms, human participants, and sound making props with a few exceptions. In each session, various clips are recorded with combinations of that session's participants acting some simple scenes and interacting between them and with the sound making props. The scenes are not strongly scripted; instead they are based on generic instructions on what kind of sound events to contain and they are otherwise improvised by the participants. The instructions serve as a rough guide to ensure adequate event activity and inclusion of events from the target sound classes in a clip. 

Similarly to the previous three challenges, the recordings are converted to two 4-channel spatial formats: first-order Ambisonics (FOA) and tetrahedral microphone array (MIC), both derived from the original 32-channel recordings. Conversion of the Eigenmike recordings to FOA following the SN3D normalization scheme (or ambiX) was performed with measurement-based filters according to \cite{politis2017comparing}. Regarding the MIC format, channels 6, 10, 26, and 22 of the Eigenmike were selected, corresponding to a nearly tetrahedral arrangement of (azimuth, elevation, radius) spherical coordinates ($\ang{45}$, $\ang{35}$, 4.2 cm), ($\ang{-45}$, $\ang{-35}$, 4.2 cm), ($\ang{135}$, $\ang{-35}$, 4.2 cm) and ($\ang{-135}$, $\ang{35}$, 4.2 cm). Analytical expressions of the directional responses of each format can be found in the DCASE2020 challenge report \cite{Politis2020}. Finally, the converted recordings were downsampled to 24kHz.

The dataset is split into a development set (\emph{dev-set}) and evaluation set (\emph{eval-set}). The development set totals about 4 hrs 52 mins, of which 70 recording clips amounting to about 2 hrs are recorded in 4 different rooms in Tokyo and 51 recordings amounting to about 3 hrs are recorded in 7 different rooms in Tampere. To aid the development process, the development set is further split into a training part (\emph{dev-set-train}, 40+27 clips in 2+4 rooms in Tokyo+Tampere) and a testing part (\emph{dev-set-test}, 30+24 clips in 2+3 rooms in Tokyo+Tampere).


\begin{table}[t]
\begin{tabular}{p{2.5cm}|p{5cm}}
\textbf{Target Class}        & \textbf{Related Audioset subclasses}                                                                                                       \\
\hline
\textit{Telephone}           & \emph{Telephone bell ringing}, \emph{Ringtone} (no musical ringtones)                                        \\
\hline
\textit{Domestic sounds}     & \emph{Vacuum cleaner, Mechanical fan, Boiling} (produced by hoover, air circulator, water boiler)\\
\hline
\textit{Door, open or close} & Combination of \emph{Door} \& \emph{Cupboard, open or close}                                                                                             \\
\hline
\textit{Music}               & \emph{Background music} \& \emph{Pop music},
(played by a loudspeaker in the room)                          \\
\hline
\textit{Musical instrument}  & \emph{Acoustic guitar, Marimba, Xylophone, Cowbell, Piano, Rattle (instrument)}                        \\
\hline
\textit{Bell}                & Combination of sounds from hotel bell and glass bell, closer to \emph{Bicycle bell} \& single \emph{Chime}
\end{tabular}
\caption{Relation of target classes to specific Audioset classes.}
\label{table:classes_audioset}
\end{table}
\begin{table*}[ht]
\begin{adjustbox}{max width=\textwidth}
\small
\begin{tabular}{l||l||l|l|l|l|l|l|l|l|l|l|l|l|l}
                                                                            & \textbf{Global} & \textbf{\begin{tabular}[c]{@{}l@{}}Fem.\\ speech\end{tabular}} & \textbf{\begin{tabular}[c]{@{}l@{}}Male \\ speech\end{tabular}} & \textbf{Clap} & \textbf{Phone} & \textbf{Laugh} & \textbf{\begin{tabular}[c]{@{}l@{}}Dom.\\ sounds\end{tabular}} & \textbf{Footsteps} & \textbf{Door} & \textbf{Music} & \textbf{\begin{tabular}[c]{@{}l@{}}Music.\\ instr.\end{tabular}} & \textbf{Faucet} & \textbf{Bell} & \textbf{Knock} \\ \hline
\begin{tabular}[c]{@{}l@{}}Frame coverage \\ (\% total frames)\end{tabular} & 84.7            & 20.4                                                             & 37.6                                                            & 0.7               & 1.4                & 2.7               & 17.9                                                               & 1.3                & 0.6           & 29.4           & 4.0                                                                   & 1.7                & 1.5           & 0.1            \\
Max. polyphony                                                              & 5               & 2                                                                & 3                                                               & 2                 & 1                  & 4                 & 1                                                                  & 1                  & 1             & 1              & 4                                                                     & 1                  & 1             & 1              \\
Mean polyphony                                                              & 1.5             & 1.04                                                             & 1.07                                                            & 1.17              & 1.00               & 1.18              & 1.00                                                               & 1.00               & 1.00          & 1.00           & 1.86                                                                  & 1.00               & 1.00          & 1.00           \\
\begin{tabular}[c]{@{}l@{}}Polyphony 1 \\ (\% active frames)\end{tabular}   & 61.5            & 96.1                                                             & 93.3                                                            & 83.4              & 100                & 84.0              & 100                                                                & 100                & 100           & 100            & 52.2                                                                  & 100                & 100           & 100            \\
Polyphony 2                                                                 & 29.55           & 3.9                                                              & 6.5                                                             & 16.6              & 0                  & 14.5              & 0                                                                  & 0                  & 0             & 0              & 16.6                                                                  & 0                  & 0             & 0              \\
Polyphony 3                                                                 & 7.15            & 0                                                                & 0.2                                                             & 0                 & 0                  & 1.1               & 0                                                                  & 0                  & 0             & 0              & 24.2                                                                  & 0                  & 0             & 0              \\
Polyphony 4                                                                 & 1.6             & 0                                                                & 0                                                               & 0                 & 0                  & 0.4               & 0                                                                  & 0                  & 0             & 0              & 7.0                                                                   & 0                  & 0             & 0              \\
Polyphony 5                                                                 & 0.2             & 0                                                                & 0                                                               & 0                 & 0                  & 0                 & 0                                                                  & 0                  & 0             & 0              & 0                                                                     & 0                  & 0             & 0             
\end{tabular}
\end{adjustbox}
\caption{Dataset class activity and polyphony information. The mean polyphony is computed over active frames only having one or more events present.}
\label{table:class_activity}
\end{table*}

\subsection{Recording setup and process}

Each scene was captured with 4 types of sensors: a) a high resolution 32-channel SMA (Eigenmike em32 by mh Acoustics\footnote{\url{https://mhacoustics.com/products#eigenmike1}}) recording the main multichannel audio for the challenge, b) a 360\degree camera (Ricoh Theta V\footnote{\url{https://theta360.com/en/about/theta/v.html}}) mounted about 10 cm above the SMA, c) a motion capture (mocap) system of infrared cameras surrounding the scene, tracking reflective markers mounted on the main actors and sound sources of interest (Optitrack Flex 13\footnote{\url{https://optitrack.com/cameras/flex-13/}}), and d) wireless microphones mounted on the same tracked actors and sound sources, providing close-miked recordings of the main sound events (R\o de Wireless Go II\footnote{\url{https://rode.com/en/microphones/wireless/wirelessgoii}}). For each recording session, a suitable position of the Eigenmike and Ricoh Theta V would be decided in order to cover the scene from a central position, while taking into account the intended scenarios and the specific room constraints. The origin of the mocap system was then set at ground level on the same position and the height of the Eigenmike was set at 1.5 m, while the mocap cameras were positioned at the boundaries of the room. Tracking markers were mounted to independent sound sources (such as next to the water sink, on a mobile phone on a table, on a hoover, or next to a guitar's soundhole). Head markers were additionally provided to the participants before each scene recording, in the form of headbands or hats. Tracking the head served as the reference point for all human made sounds. Mouth position for \emph{speech} and \emph{laughter} sounds, feet stepping position for \emph{footstep} sounds, and hand position for \emph{clapping} sounds were each approximated with a fixed translation from the head-tracking center close to the top of the head. Regarding clapping, participants were instructed to clap about 20 cm in front of their face to improve the position approximation. Head rotations were also logged during the scene with respect to the global coordinate frame of the mocap system. Finally, the wireless microphones were mounted to the lapel of each actor and to additional independent sound sources being at a distance from the actors. 

Recording would start on all devices before the beginning of a scene and would stop right after. A clapper sound would initiate the acting and it would serve as a reference signal for synchronization between the different types of recordings, including the mocap system which could record a monophonic audio side-signal for that exact reason. All 4 types of recordings were manually synchronized based on the clapper sound and subsequently cropped and stored at the end of each recording session.

\subsection{Annotation process}

Spatiotemporal annotations of the sound events were conducted manually by the authors and research assistants. Three types of information were required in order to obtain such annotations: a) the subset of the target classes that were active in each scene, b) the temporal activity of such class instances, and c) the position of each such instance when active. (a) was observed and logged during each scene recording. (b) was manually annotated by listening to the wireless microphone recordings. Since each such microphone would capture prominently sounds produced by the human actor or source it was assigned to, onset, offsets, source, and class information of each event could be conveniently extracted. In scenes or instances where associating an event to a source was ambiguous purely by listening, annotators would consult the video recordings to establish the correct association. The temporal annotation resolution was set to 100 msec.

After onset, offset, and class information of events was established for each source and actor in the scene, the positional annotations (c) were extracted for each such event by masking the tracker data with the temporal activity window of the event. Additionally, class-specific translations to the tracking data were applied if necessary, as mentioned earlier for most human made sounds. Positional information was logged in Cartesian coordinates with respect to the mocap system's origin. Since the dataset targets directional localization instead of absolute position estimation, all tracked positions were converted to directions-of-arrival with respect to the center of the Eigenmike. Finally, the class, temporal, and spatial annotations were combined and converted to the text format used in the previous DCASE2019-2021 challenges. Validation of the annotations was performed by observing and listening to the 360\degree videos form Ricoh Theta V, overlapped with generated videos of the event activities visualized as labeled markers positioned at their respective DOAs on the 360\degree video plane.

\subsection{Target sound classes}

A set of 13 target sound classes are selected to be annotated, based on the sound events captured adequately in the recorded scenes. The class labels are selected to conform to the Audioset ontology \cite{gemmeke2017audio} and they are: \textbf{\emph{female speech/woman speaking, male speech/man speaking, clapping, telephone, laughter, domestic sounds, walk/footsteps, door open or close, music, musical instrument, water tap/faucet, bell, knock}}. The speech class contains speech in a few different languages. Since some of these labels correspond to super classes with a large diversity of sounds and number of subclasses in the ontology (e.g. \emph{domestic sounds} or \emph{musical instrument}) we provide some additional information on the subset of sounds encountered in the recordings for some of the target classes, in the form of more specific audioset-related labels. That information can aid training and testing of systems; however, only the more general target labels are provided as annotations. This information is summarized in Table~\ref{table:classes_audioset}. Target classes not included in the table have an one-to-one relationship with the similarly named Audioset ones. Apart from the sound events belonging to one of the target classes, additional directional sound events occur in the recordings which are not annotated and are treated as directional interferers; examples include \emph{computer keyboard}, \emph{shuffling cards}, and \emph{dishes, pots, and pans}. Additionally, there is natural background noise in all recordings, mostly HVAC-related, ranging from low to considerable levels. It is expected that such background noise would be distinguishable from the noisy target sources such as vacuum cleaner or mechanical fan, since contrary to those sources it manifests as diffuse or weakly-directional sound. Information on the percentage of frames that each class is active, and the degree of polyphony for each class and globally, based on the annotations, is presented in Table~\ref{table:class_activity}.

\section{Baseline}
\label{sec:baseline}

\subsection{Model architecture}

The baseline for this year's challenge\footnote{\url{https://github.com/sharathadavanne/seld-dcase2022}} is similar to the one used in DCASE2021, with one major difference. It is based on a convolutional recurrent neural network (CRNN) stemming from the original SELDnet architecture \cite{adavanne2018sound} proposed in the first challenge, but improved with the \emph{activity-coupled Cartesian direction of arrival} output representation (ACCDOA) \cite{shimada2021accdoa}. ACCDOA increases SELD performance using a single homogeneous regression loss instead of the original's combination of a classification cross-entropy loss and localization regression loss. One limitation of the original ACCDOA proposal and the DCASE2021 baseline is the inability of the model to handle multiple events of the same class occurring simultaneously. To handle this case, the current baseline adopts the strategy of \emph{multi-ACCDOA} (mACCDOA) proposed by \cite{shimada2022multi} with the output of the model switched to a track-based format corresponding to a maximum number of simultaneous events, instead of the previous purely class-based format. Hence, the network receives a sequence of $T$ STFT frames of multichannel features and instead of the ACCDOA model outputting $T/5\times C \times 3$ Cartesian vector coordinates indicating the DoA (encoded in the vector direction) and activity (encoded in the vector magnitude) of each class, the mACCDOA model outputs $T/5\times N\times C \times 3$ vector coordinates, where $C$ is the number of target classes and $N$ the maximum assumed number of co-occuring events in the recordings. For the current baseline $N$ is set to 3 maximum simultaneous sources, while a value of 0.5 is used as the threshold on the length of the output vectors to indicate track and class activity. Note that a reduction of the STFT temporal resolution by a factor of 5 is performed to match the resolution of the annotations at every 100 msec.

Similarly to the previous years, the model receives different inputs depending on the recording format. Four-channel spectrograms for both formats are computed with 1024-point FFTs using a 40 msec hanning window and 20 msec hop length at 24kHz. Log-mel spectrograms are additionally extracted from the STFT ones, for both MIC and FOA formats, at 64 mel-bands. Spatial features in the form of acoustic intensity vectors for each STFT bin are computed from the FOA spectrograms and aggregated into a similar number of mel-bands. While for the MIC format, 6 generalized cross-correlation (GCC) sequences are computed for every frame and truncated to the same number of lag values as the mel-bands following \cite{Cao2019}. The 4-channel mel spectrograms are stacked along with the respective spatial features for each format across the channel dimension, resulting in $(4+3)\times 64$ features for the FOA format and $(4+6)\times 64$ features for the MIC format, for every input frame. In this year's baseline we additionally include the option of the \emph{SALSA-lite} spatial features for the MIC format \cite{nguyen2022salsa}. These features constitute essentially frequency-normalized inter-channel phase differences between a reference microphone and the rest and, contrary to the GCC, they have the advantage of being spectrotemporally aligned with the spectrograms with increased robustness in multi-source scenarios. The baseline implementation avoids mel-band conversion; instead the original STFT spectrograms and the respective spatial features are truncated to include bins up to about 9 kHz, following \cite{nguyen2022salsa}. Hence, the size of input features are $(4+3)\times 382$ in this case, for every input frame. 

\begin{figure}[t]
  \centering
  \centerline{\includegraphics[width=0.4\textwidth ,keepaspectratio]{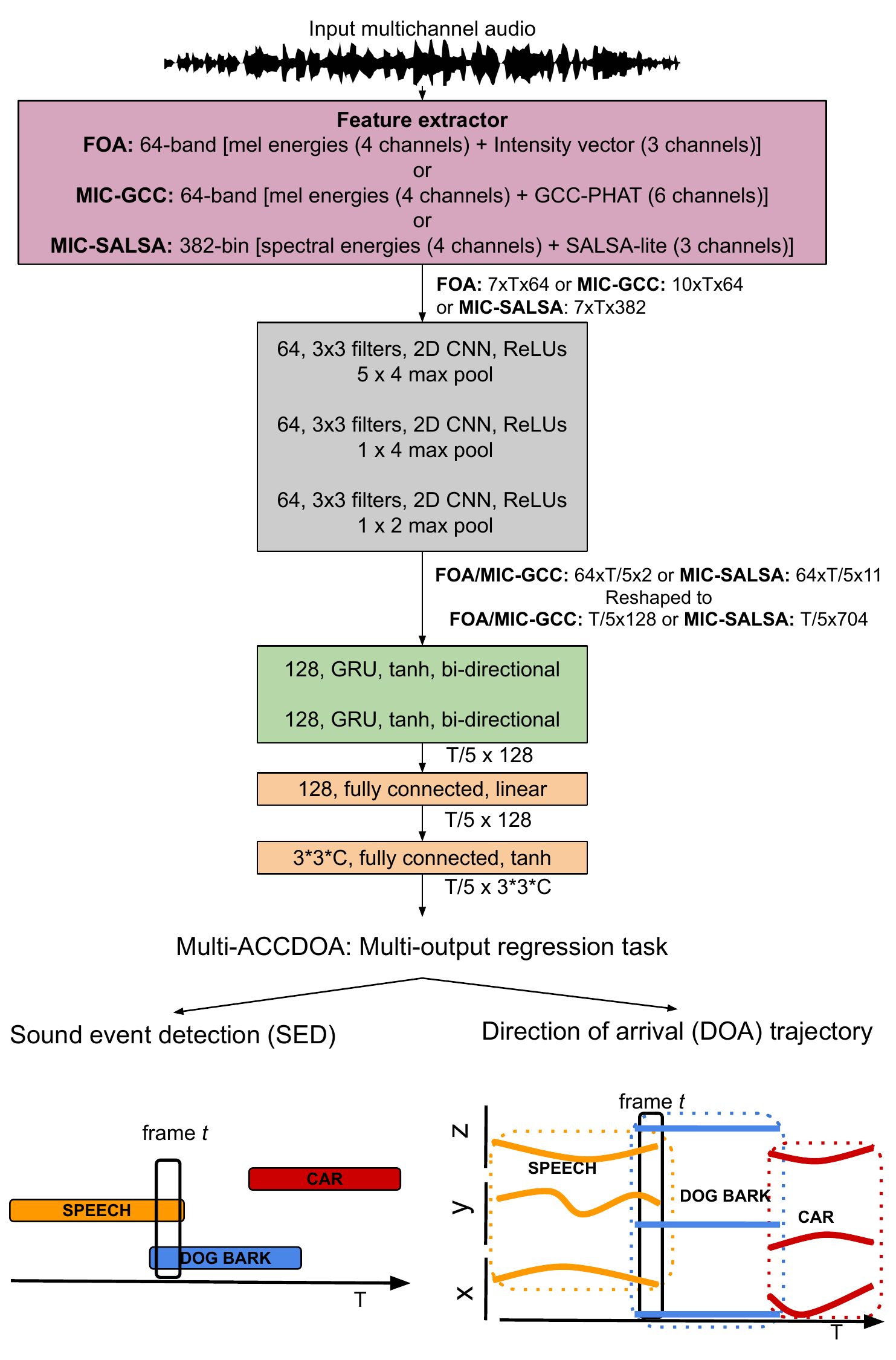}} 
  \caption{Convolutional recurrent neural network with ACCDOA loss for SELD.}
  \label{fig:crnn}
\end{figure}

\subsection{Model training}
\label{sec:model_training}

The baseline model is trained and evaluated twice: firstly on the development set only to report initial baseline results for the participants to compare against during development, secondly it is trained on the development set and tested on the evaluation set, with results reported after the completion of the evaluation phase of the challenge. Since, the amount of training material may be insufficient for the complexity of the task, additional material is synthesized during training. Those synthetic mixtures are generated with the same generation process and the same spatial room impulse responses as the TAU-NIGENS Spatial Sound Events 2020-2021 datasets used in the development and evaluation of DCASE2020-2021 challenge. 1200 one-minute spatial mixtures are synthesized (\emph{synth-set}) using SRIRs from 9 rooms in TAU and sound event samples sourced from FSD50K \cite{fonseca2021fsd50k}. The samples are selected on the basis of their annotated labels which follow the Audioset ontology. The synthetic mixtures are made publicly available for reproducibility\footnote{\url{https://doi.org/10.5281/zenodo.6406873}} along with the list of the selected FSD50K sound samples. Additionally, the SRIRs are also publicly shared\footnote{\url{https://doi.org/10.5281/zenodo.6408611}} along with the scene generation code\footnote{\url{https://github.com/danielkrause/DCASE2022-data-generator}}, so that participants can generate their own synthetic mixtures for training following the same process if desired. The sets and splits for training and testing of the baseline for each phase are summarized in Table~\ref{tab:baseline_train}.

\begin{table}[ht]
\begin{adjustbox}{max width=\columnwidth}
\begin{tabular}{l|ll}
\textbf{Phase} & \textbf{Training} & \textbf{Testing} \\ \hline
\textbf{Development}  & synth-set + dev-set-train & dev-set-test \\
\textbf{Evaluation} & synth-set + dev-set-train + dev-set-test & eval-set  \\
\end{tabular}
\end{adjustbox}
\caption{Datasets and splits used for baseline training for results on the development set and on the evaluation set.}
\label{tab:baseline_train}
\end{table}

\section{Evaluation}
\label{sec:evaluation}

The development dataset (both \emph{dev-set-train} and \emph{dev-set-test}) is published to the challenge participants at the commencement of the challenge, including annotations, while the evaluation dataset is made public after the completion of the development phase and at the commencement of the evaluation phase, with annotations withheld by the organizers. At the end of the evaluation phase participants submit their system outputs on the evaluation dataset and the organizers perform the evaluation.
Additionally, participants are required to report results on the development dataset, following the provided train-test split for a consistent comparison with the baseline results and with the other submissions. Note that contrary to the previous challenges, participants are allowed to use external data during training, such as sample banks of sound events, SRIRS, spatial background noise recordings, pre-trained networks, and others, in order to generate additional training material covering more diverse conditions than the provided dataset. Generating the \emph{synth-set} dataset and using it to improve the baseline performance constitutes just one such example of external data usage.


\subsection{Evaluation metrics}

The submissions are evaluated with the joint localization-detection metrics studied in \cite{mesaros2019joint, politis2020overview} and introduced first-time in DCASE2020. A brief description of the metrics follows. The first two metrics, the location-dependent error rate ($ER_X$) and F1-score ($F_X$) for a spatial threshold $X$ are based on true positives ($TP$), false negatives ($FN$), and false positives ($FP$) determined not only by correct, missed, or wrong detections, but also based on detections being closer or further than a distance threshold $X$ from the reference. In the present case of DOA estimation the threhold is angular, and it is taken to be $X = 20^\circ$. For each class $c\in[1,...,C]$ detections are computed in a segment-based fashion \cite{mesaros2016metrics} in 1 second segments. For each segment $P_c$ predicted events of class $c$ are associated with $R_c$ reference events of the same class. False negatives and false positive are counted for missed or extraneous detections respectively
\begin{align}
    FN_c &= \max(0, R_c-P_c)\\
    FP_c^{(d)} &= \max(0, P_c-R_c)
\end{align}
where the $(d)$ superscript indicates purely detection based false positives to differentiate from the spatial ones. Furthermore, 
$TP_c=\min(P_c,R_c)$ predictions are spatially associated with references using the Hungarian algorithm, which can also be considered as the unthreholded true positives. Then the spatial threshold is applied to those associated predictions, which moves $FP_{c,\geq 20^\circ} \leq TP_c$ predictions further than the threshold from true positives to spatial false positives. The combined number of false positives and the remaining matched true positives per class are
\begin{align}
    FP_c &= FP_c^{(d)} + FP_{c,\geq 20^\circ}\\
    TP_{c,\leq 20^\circ} &= TP_c - FP_{c,\geq 20^\circ}
\end{align}
Based on $FN_c, FP_c$ and $TP_{c,\leq 20^\circ}$ we form the location-dependent error rate $ER_{20^\circ}$ and F1-score $F_{20^\circ}$. Contrary to the previous challenges, in which $F_{20^\circ}$ was micro-averaged, in this challenge evaluation is based on macro-averaging of F1-score, with $F_{20^\circ} = \sum_c F_{c,20^\circ}/C$.

Localization accuracy is additionally evaluated through a class-dependent localization error $LE_c$, computed as the mean angular error of the spatially associated predictions per class (for $TP_c\neq0$), and a localization recall $LR_c$
\begin{align}
    LE_c &= \sum_k \theta_k/TP_c\\
    LR_c &= TP_c/(TP_c + FN_c)
\end{align}
with $\theta_k$ being the angular error between the $k$th matched prediction and reference. Both $LE_c$ and $LR_c$ are averaged across all frames that have any true positives or any references, respectively, and then macro-averaged $LE_{CD} = \sum_c LE_c/C$ and $LR_{CD} = \sum_c LR_c/C$. Note that the localization error and recall are not spatially thresholded in order to give more varied complementary information to the location-dependent F1-score, presenting localization accuracy beyond the spatial threshold. Note that all the metrics above treat detections on the instance level of each class to cope with multiple simultaneous reference events of the same class occurring, for example, at different locations. For more details the reader is referred to \cite{politis2020overview}. 

\subsection{Results}
\label{sec:Results}

\begin{table}[t]
\begin{adjustbox}{max width=\columnwidth}
\begin{tabular}{lccccc}
                                  & \multicolumn{1}{l}{$ER_{20^\circ{}}\downarrow$} & \multicolumn{1}{l}{$F_{20^\circ{}}\uparrow$} & \multicolumn{1}{l}{$F_{20^\circ{}}\uparrow$} & \multicolumn{1}{l}{$LE_{CD}\downarrow$} & \multicolumn{1}{l}{$LR_{CD}\uparrow$} \\
                                  &  & (macro) & (micro) &  &  \\                                  
\hline
\multicolumn{6}{l}{\textbf{Development set}}                                                                                                                                                                                          \\ \hline
\multicolumn{1}{l|}{\textbf{FOA}} & 0.71                                & 0.21                                       & 0.36                                       & 29.3$^\circ$                           & 0.46                            \\
\multicolumn{1}{l|}{\textbf{MIC}} & 0.71                                & 0.18                                       & 0.36                                       & 32.2$^\circ$                           & 0.47                            \\ \hline
\end{tabular}
\end{adjustbox}
\caption{Baseline results on the \emph{dev-set-test} split of the development set.}
\label{table:baseline_results}
\end{table}

Results of the baseline on the development set are presented on Table~\ref{table:baseline_results}, for both FOA and MIC formats. The baseline was trained as indicated in Sec.~\ref{sec:model_training} using the additional synthetic spatial mixtures of \emph{synth-set}. It is noted that the SRIRs used for the generation of those mixtures were captured in TAU spaces that were different than the ones were the scene recordings of the STARS22 dataset occurred. Two training strategies were tested with regards to incorporating the synthetic data. The first was based on initial training of the model on the synthetic data, followed by fine-tuning with the \emph{dev-set-train} split of the development set. The second simply mixed both the \emph{synth-set} and the \emph{dev-set-train} and trained with the combined dataset. Better results were obtained with the mixed strategy and these are the ones presented here. Regarding the MIC format, both the GCC features and the SALSA-lite features were tested. Slightly better results were obtained with the GCC features and reported here. That may be attributed to the fact that even though the SALSA-lite features show a clear advantage for densely populated multi-source scenes \cite{nguyen2022salsa} such as the ones in DCASE2021 dataset, for more sparse scenes as the ones in STARSS22 that advantage may be diminished. Finally, both the micro and macro versions of the F1-score are presented here, with a clear drop in performance in the macro version, as expected with a dataset of such unbalanced presence of target classes (evident in Table~\ref{table:class_activity}).

\section{Conclusions}
\label{sec:conclusions}

This report presents the specifications of the STARS22 dataset, which consists of spatial recordings of real scenes with annotations of sound events of target classes both spatially and temporally. The dataset allows evaluation of SELD systems in scene recordings in more challenging real conditions with a natural composition of sound events. Additionally, it opens some novel possibilities for acoustic scene analysis and machine listening not possible with available synthetic datasets. The dataset serves as the development and evaluation dataset of the SELD challenge of DCASE2022, and is accompanied by a baseline similar to the previous iterations, with the exception of handling multiple simultaneous instances of the same class through the multi-ACCDOA representation and support for additional input formats. Results on the development dataset indicate that with use of external data and a suitable training strategy the baseline can achieve a reasonable performance on the new dataset.

\section{Acknowledgment}
\label{sec:acknowledgment}

The dataset collection and annotation at Tampere University has been funded by Google.

\bibliographystyle{IEEEtran}
\bibliography{refs}

%
%
%
%
%
%
%
%
%

\end{sloppy}
\end{document}